% This is AA.CMM, the plain TeX macro package
% (CM version) from Springer-Verlag
% for the Astronomy and Astrophysics Main Journal
% Version 2.0 as of 25 Feb 1991
%
% Test for recursive or multiple loading of Springer macro packages
\def\SpringerMacroPackageNameATest{AA}%
\let\next\relax
\ifx\SpringerMacroPackageNameA\undefined
  \message{Loading the \SpringerMacroPackageNameATest\space
           macro package from Springer-Verlag...}%
\else
  \ifx\SpringerMacroPackageNameA\SpringerMacroPackageNameATest
    \message{\SpringerMacroPackageNameA\space macro package
             from Springer-Verlag already loaded.}%
    \let\next\endinput
  \else
    \message{DANGER: \SpringerMacroPackageNameA\space from
             Springer-Verlag already loaded, will try to proceed.}%
  \fi
\fi
\next
\def\SpringerMacroPackageNameA{AA}%
% now call all the sub-macros
% indention of equations
\newskip\mathindent      \mathindent=0pt
% \titlea
\newskip\tabefore \tabefore=20dd plus 10pt minus 5pt      % space above
\newskip\taafter  \taafter=10dd                           % space below
% \titleb
\newskip\tbbeforeback    \tbbeforeback=-20dd              % corrective space to a \titlea
\newskip\tbbefore        \tbbefore=17pt plus 7pt minus3pt % spaceabove
\newskip\tbafter         \tbafter=8pt                     % space below
% \titlec
\newskip\tcbeforeback    \tcbeforeback=-3pt               % corrective space to a \titleb
\advance\tcbeforeback by -10dd                            % corrective space to a \titleb
\newskip\tcbefore        \tcbefore=10dd plus 5pt minus 1pt% space above
\newskip\tcafter         \tcafter=6pt                     % space below
% \titled
\newskip\tdbeforeback    \tdbeforeback=-3pt                  % corrective space to a \titlec
\advance\tdbeforeback by -10dd                               % corrective space to a \titlec
\newskip\tdbefore        \tdbefore=10dd plus 4pt minus 1pt   % space above
% \petit
\newskip\petitsurround
\petitsurround=6pt\relax
% \ack
\newskip\ackbefore      \ackbefore=10dd plus 5pt             % space above
\newskip\ackafter       \ackafter=6pt                        % space below
% indention of lists
\newdimen\itemindent    \newdimen\itemitemindent
\itemindent=1.5em       \itemitemindent=2\itemindent
 \font \tatt            = cmbx10 scaled \magstep3
 \font \tats            = cmbx10 scaled \magstep1
 \font \tamt            = cmmib10 scaled \magstep3
 \font \tams            = cmmib10 scaled \magstep1
 \font \tamss           = cmmib10
 \font \tast            = cmsy10 scaled \magstep3
 \font \tass            = cmsy10 scaled \magstep1
 \font \tbtt            = cmbx10 scaled \magstep2
 \font \tbmt            = cmmib10 scaled \magstep2
 \font \tbst            = cmsy10 scaled \magstep2
\catcode`@=11    % use @ as a normal character
\vsize=23.5truecm
\hoffset=-1true cm
\voffset=-1true cm
\normallineskip=1dd
\normallineskiplimit=0dd
\newskip\ttglue%
\def\ifundefin@d#1#2{%
\expandafter\ifx\csname#1#2\endcsname\relax}
\def\getf@nt#1#2#3#4{%
\ifundefin@d{#1}{#2}%
\global\expandafter\font\csname#1#2\endcsname=#3#4%
\fi\relax
}
\newfam\sffam
\newfam\scfam
\def\makesize#1#2#3#4#5#6#7{%
 \getf@nt{rm}{#1}{cmr}{#2}%
 \getf@nt{rm}{#3}{cmr}{#4}%
 \getf@nt{rm}{#5}{cmr}{#6}%
 \getf@nt{mi}{#1}{cmmi}{#2}%
 \getf@nt{mi}{#3}{cmmi}{#4}%
 \getf@nt{mi}{#5}{cmmi}{#6}%
 \getf@nt{sy}{#1}{cmsy}{#2}%
 \getf@nt{sy}{#3}{cmsy}{#4}%
 \getf@nt{sy}{#5}{cmsy}{#6}%
 \skewchar\csname mi#1\endcsname ='177
 \skewchar\csname mi#3\endcsname ='177
 \skewchar\csname mi#5\endcsname ='177
 \skewchar\csname sy#1\endcsname ='60
 \skewchar\csname sy#3\endcsname='60
 \skewchar\csname sy#5\endcsname='60
\expandafter\def\csname#1size\endcsname{%
 \normalbaselineskip=#7
 \normalbaselines
 \setbox\strutbox=\hbox{\vrule height0.75\normalbaselineskip%
    depth0.25\normalbaselineskip width0pt}%
 \textfont0=\csname rm#1\endcsname
 \scriptfont0=\csname rm#3\endcsname
 \scriptscriptfont0=\csname rm#5\endcsname
    \def\oldstyle{\fam1\csname mi#1\endcsname}%
 \textfont1=\csname mi#1\endcsname
 \scriptfont1=\csname mi#3\endcsname
 \scriptscriptfont1=\csname mi#5\endcsname
 \textfont2=\csname sy#1\endcsname
 \scriptfont2=\csname sy#3\endcsname
 \scriptscriptfont2=\csname sy#5\endcsname
 \textfont3=\tenex\scriptfont3=\tenex\scriptscriptfont3=\tenex
   \def\rm{%
 \fam0\csname rm#1\endcsname%
   }%
   \def\it{%
 \getf@nt{it}{#1}{cmti}{#2}%
 \textfont\itfam=\csname it#1\endcsname
 \fam\itfam\csname it#1\endcsname
   }%
   \def\sl{%
 \getf@nt{sl}{#1}{cmsl}{#2}%
 \textfont\slfam=\csname sl#1\endcsname
 \fam\slfam\csname sl#1\endcsname}%
   \def\bf{%
 \getf@nt{bf}{#1}{cmbx}{#2}%
 \getf@nt{bf}{#3}{cmbx}{#4}%
 \getf@nt{bf}{#5}{cmbx}{#6}%
 \textfont\bffam=\csname bf#1\endcsname
 \scriptfont\bffam=\csname bf#3\endcsname
 \scriptscriptfont\bffam=\csname bf#5\endcsname
 \fam\bffam\csname bf#1\endcsname}%
   \def\tt{%
 \getf@nt{tt}{#1}{cmtt}{#2}%
 \textfont\ttfam=\csname tt#1\endcsname
 \fam\ttfam\csname tt#1\endcsname
 \ttglue=.5em plus.25em minus.15em
   }%
  \def\sf{%
\getf@nt{sf}{#1}{cmss}{10 at #2pt}%
\textfont\sffam=\csname sf#1\endcsname
\fam\sffam\csname sf#1\endcsname}%
   \def\sc{%
 \getf@nt{sc}{#1}{cmcsc}{10 at #2pt}%
 \textfont\scfam=\csname sc#1\endcsname
 \fam\scfam\csname sc#1\endcsname}%
\rm }}
\makesize{IXf}{9}{VIf}{6}{Vf}{5}{10.00dd}
\def\normalsize{\IXfsize
\def\sf{%
   \getf@nt{sf}{IXf}{cmss}{9}%
   \getf@nt{sf}{VIf}{cmss}{10 at 6pt}%
   \getf@nt{sf}{Vf}{cmss}{10 at 5pt}%
   \textfont\sffam=\csname sfIXf\endcsname
   \scriptfont\sffam=\csname sfVIf\endcsname
   \scriptscriptfont\sffam=\csname sfVf\endcsname
   \fam\sffam\csname sfIXf\endcsname}%
}%
\newfam\mibfam
\def\mib{%
   \getf@nt{mib}{IXf}{cmmib}{10}%
   \getf@nt{mib}{VIf}{cmmib}{10 at6pt}%
   \getf@nt{mib}{Vf}{cmmib}{10 at5pt}%
   \textfont\mibfam=\csname mibIXf\endcsname
   \scriptfont\mibfam=\csname mibVIf\endcsname
   \scriptscriptfont\mibfam=\csname mibVf\endcsname
   \fam\mibfam\csname mibIXf\endcsname}%
\makesize{Xf}{10}{VIf}{6}{Vf}{5}{10.00dd}
\Xfsize
\it\bf\tt\rm

\def\tentt{\ttXf}

\normalsize
\it\bf\tt\sf\mib\rm
\def\boldmath{\textfont1=\mibIXf \scriptfont1=\mibVIf
\scriptscriptfont1=\mibVf}
\newdimen\fullhsize
\newcount\verybad \verybad=1010
\let\lr=L%
\fullhsize=40cc
\hsize=19.5cc
\def\fullline{\hbox to\fullhsize}
\def\makefootline{\baselineskip=10dd \fullline{\the\footline}}
\def\makeheadline{\vbox to 0pt{\vskip-22.5pt
            \fullline{\vbox to 8.5pt{}\the\headline}\vss}\nointerlineskip}
\hfuzz=2pt
\vfuzz=2pt
\tolerance=1000
\abovedisplayskip=3 mm plus6pt minus 4pt
\belowdisplayskip=3 mm plus6pt minus 4pt
\abovedisplayshortskip=0mm plus6pt
\belowdisplayshortskip=2 mm plus4pt minus 4pt
\parindent=1.5em
\newdimen\stdparindent\stdparindent\parindent
\frenchspacing
\nopagenumbers
\predisplaypenalty=600        % Make a page break before a display harder
\displaywidowpenalty=2000     % and even harder for a widow display.
\def\widowsandclubs#1{\global\verybad=#1
   \global\widowpenalty=\the\verybad1      % default: 10101
   \global\clubpenalty=\the\verybad2  }    % default: 10102
\widowsandclubs{1010}
\def\paglay{\headline={{\normalsize\hsize=.75\fullhsize\ifnum\pageno=1
\vbox{\hrule\line{\vrule\kern3pt\vbox{\kern3pt
\hbox{\bf A\&A manuscript no.}
\hbox{(will be inserted by hand later)}
\kern3pt\hrule\kern3pt
\hbox{\bf Your thesaurus codes are:}
\hbox{\rightskip=0pt plus3em\advance\hsize by-7pt
\vbox{\bf\noindent\ignorespaces\the\THESAURUS}}
\kern3pt}\hfil\kern3pt\vrule}\hrule}
\rlap{\quad\AALogo}\hfil
\else\normalsize\ifodd\pageno\hfil\folio\else\folio\hfil\fi\fi}}}
\makesize{VIIIf}{8}{VIf}{6}{Vf}{5}{9.00dd}
      \getf@nt{sf}{VIIIf}{cmss}{8}%
      \getf@nt{sf}{VIf}{cmss}{10 at 6pt}%
      \getf@nt{sf}{Vf}{cmss}{10 at 5pt}%
      \getf@nt{mib}{VIIIf}{cmmib}{10 at 8pt}%
      \getf@nt{mib}{VIf}{cmmib}{10 at 6pt}%
      \getf@nt{mib}{Vf}{cmmib}{10 at 5pt}%
\VIIIfsize\it\bf\tt\rm
\normalsize
\def\petit{\VIIIfsize
   \def\sf{%
      \getf@nt{sf}{VIIIf}{cmss}{8}%
      \getf@nt{sf}{VIf}{cmss}{10 at 6pt}%
      \getf@nt{sf}{Vf}{cmss}{10 at 5pt}%
      \textfont\sffam=\csname sfVIIIf\endcsname
      \scriptfont\sffam=\csname sfVIf\endcsname
      \scriptscriptfont\sffam=\csname sfVf\endcsname
      \fam\sffam\csname sfVIIIf\endcsname
}%
\def\mib{%
   \getf@nt{mib}{VIIIf}{cmmib}{10 at 8pt}%
   \getf@nt{mib}{VIf}{cmmib}{10 at 6pt}%
   \getf@nt{mib}{Vf}{cmmib}{10 at 5pt}%
   \textfont\mibfam=\csname mibVIIIf\endcsname
   \scriptfont\mibfam=\csname mibVIf\endcsname
   \scriptscriptfont\mibfam=\csname mibVf\endcsname
   \fam\mibfam\csname mibIXf\endcsname}%
\def\boldmath{\textfont1=\mibVIIIf\scriptfont1=\mibVIf
\scriptscriptfont1=\mibVf}%
\let\bfIXf=\bfVIIIf
 \if Y\REFEREE \normalbaselineskip=2\normalbaselineskip
 \normallineskip=2\normallineskip\fi
 \setbox\strutbox=\hbox{\vrule height7pt depth2pt width0pt}%
 \normalbaselines\rm}%
\def\begpet{\vskip\petitsurround
\bgroup\petit}%  Beginn eines Paragraphen in petit
\def\endpet{\vskip\petitsurround
\egroup}%  Ende eines Paragraphen in petit

 \let  \tatss           = \bfXf
 \let  \tasss           = \syXf
 \let  \tbts            = \bfXf
 \let  \tbtss           = \bfVIIIf
 \let  \tbms            = \tamss
 \let  \tbmss           = \mibVIIIf
 \let  \tbss            = \syXf
 \let  \tbsss           = \syVIIIf
\def\newline{\hfill\break}% makes a new line in the text :)
\def\rahmen#1{\vbox{\hrule\line{\vrule\vbox to#1true
cm{\vfil}\hfil\vrule}\vfil\hrule}}
\let\ts=\thinspace
\def\,{\relax\ifmmode\mskip\thinmuskip\else\thinspace\fi}
\def\unvskip{%
   \ifvmode
      \ifdim\lastskip=0pt
      \else
         \vskip-\lastskip
      \fi
   \fi}
\newtoks\eq\newtoks\eqn
\newdimen\mathhsize
\def\calcmathhsize{\mathhsize=\hsize
\advance\mathhsize by-\mathindent}
\calcmathhsize
\def\eqalign#1{\null\vcenter{\openup\jot\m@th
  \ialign{\strut\hfil$\displaystyle{##}$&$\displaystyle{{}##}$\hfil
      \crcr#1\crcr}}}
\def\displaylines#1{{}$\displ@y
\hbox{\vbox{\halign{$\@lign\hfil\displaystyle##\hfil$\crcr
    #1\crcr}}}${}}
\def\eqalignno#1{{}$\displ@y
  \hbox{\vbox{\halign
to\mathhsize{\hfil$\@lign\displaystyle{##}$\tabskip\z@skip
    &$\@lign\displaystyle{{}##}$\hfil\tabskip\centering
    &\llap{$\@lign##$}\tabskip\z@skip\crcr
    #1\crcr}}}${}}
\def\leqalignno#1{{}$\displ@y
\hbox{\vbox{\halign
to\mathhsize{\qquad\hfil$\@lign\displaystyle{##}$\tabskip\z@skip
    &$\@lign\displaystyle{{}##}$\hfil\tabskip\centering
    &\kern-\mathhsize\rlap{$\@lign##$}\tabskip\hsize\crcr
    #1\crcr}}}${}}
\def\generaldisplay{%
\ifeqno
       \ifleqno\leftline{$\displaystyle\the\eqn\quad\the\eq$}%
       \else\noindent\kern\mathindent\hbox to\mathhsize{$\displaystyle
             \the\eq\hfill\the\eqn$}%
       \fi
\else
       \kern\mathindent
       \hbox to\mathhsize{$\displaystyle\the\eq$\hss}%
\fi
\global\eq={}\global\eqn={}}%
\newif\ifeqno\newif\ifleqno
\everydisplay{\displaysetup}
\def\displaysetup#1$${\displaytest#1\eqno\eqno\displaytest}
% look for equation numbers
\def\displaytest#1\eqno#2\eqno#3\displaytest{%
\if!#3!\ldisplaytest#1\leqno\leqno\ldisplaytest
\else\eqnotrue\leqnofalse\eqn={#2}\eq={#1}\fi
\generaldisplay$$}
\def\ldisplaytest#1\leqno#2\leqno#3\ldisplaytest{\eq={#1}%
\if!#3!\eqnofalse\else\eqnotrue\leqnotrue\eqn={#2}\fi}
\newcount\eqnum\eqnum=0% register
\def\autnum{\global\advance\eqnum by 1\relax{\rm(\the\eqnum)}}
\newdimen\lindent
\lindent=\stdparindent

\def\litemitem{\par\noindent\hbox to\lindent{\hfil}%
               \hangindent=2\lindent\ltextindent}
\def\ltextindent#1{\hbox to\lindent{#1\hss}\ignorespaces}
\def\set@item@mark#1{\llap{#1\enspace}\ignorespaces}
\ifx\undefined\mathhsize
   \def\item{\par\noindent
   \hangindent\itemindent\hangafter=0
   \set@item@mark}
   \def\itemitem{\par\noindent\advance\mathhsize by-\itemitemindent
   \hangindent\itemitemindent\hangafter=0
   \set@item@mark}
\else
   \def\item{\par\noindent\advance\mathhsize by-\itemindent
   \hangindent\itemindent\hangafter=0
   \everypar={\global\mathhsize=\hsize
   \global\advance\mathhsize by-\mathindent
   \global\everypar={}}\set@item@mark}
   \def\itemitem{\par\noindent\advance\mathhsize by-\itemitemindent
   \hangindent\itemitemindent\hangafter=0
   \everypar={\global\mathhsize=\hsize
   \global\advance\mathhsize by-\mathindent
   \global\everypar={}}\set@item@mark}
\fi
\newcount\the@end \global\the@end=0
\newbox\springer@macro \setbox\springer@macro=\vbox{}
\def\typeset{\setbox\springer@macro=\vbox{\begpet\noindent
   This article was processed by the author using
   Sprin\-ger-Ver\-lag \TeX{} A\&A macro package 1991.\par
   \egroup}\global\the@end=1}
\outer\def\bye{\bigskip\typeset
\sterne=1\ifx\speciali\undefined
\else
  \loop\smallskip\noindent special character No\number\sterne:
    \csname special\romannumeral\sterne\endcsname
    \advance\sterne by 1\relax
    \ifnum\sterne<11\relax
  \repeat
\fi
\if R\lr\null\fi\vfill\supereject\end}
\def\AALogo{\setbox254=\hbox{ ASTROPHYSICS }%
\vbox{\baselineskip=10dd\hrule\hbox{\vrule\vbox{\kern3pt
\hbox to\wd254{\hfil ASTRONOMY\hfil}
\hbox to\wd254{\hfil AND\hfil}\copy254
\hbox to\wd254{\hfil\number\day.\number\month.\number\year\hfil}
\kern3pt}\vrule}\hrule}}
\def\figure#1#2{\medskip\noindent{\petit{\bf Fig.\ts#1.\
}\ignorespaces#2\par}}
\expandafter \newcount \csname c@Tl\endcsname
    \csname c@Tl\endcsname=0
\expandafter \newcount \csname c@Tm\endcsname
    \csname c@Tm\endcsname=0
\expandafter \newcount \csname c@Tn\endcsname
    \csname c@Tn\endcsname=0
\expandafter \newcount \csname c@To\endcsname
    \csname c@To\endcsname=0
\expandafter \newcount \csname c@Tp\endcsname
    \csname c@Tp\endcsname=0
\expandafter \newcount \csname c@fn\endcsname
    \csname c@fn\endcsname=0
\def \stepc#1    {\global
    \expandafter
    \advance
    \csname c@#1\endcsname by 1}
\def \resetcount#1    {\global
    \csname c@#1\endcsname=0}
\def\@nameuse#1{\csname #1\endcsname}
\def\arabic#1{\@arabic{\@nameuse{c@#1}}}
\def\@arabic#1{\ifnum #1>0 \number #1\fi}
 \def \aTa  { \goodbreak
     \bgroup
     \par
 \textfont0=\tatt \scriptfont0=\tats \scriptscriptfont0=\tatss
 \textfont1=\tamt \scriptfont1=\tams \scriptscriptfont1=\tamss
 \textfont2=\tast \scriptfont2=\tass \scriptscriptfont2=\tasss
     \baselineskip=17dd\lineskiplimit=0pt\lineskip=0pt
     \rightskip=0pt plus4cm
     \pretolerance=10000
     \noindent
     \tatt}
 \def \eTa{\vskip10pt\egroup
     \noindent
     \ignorespaces}
 \def \aTb{\goodbreak
     \bgroup
     \par
 \textfont0=\tbtt \scriptfont0=\tbts \scriptscriptfont0=\tbtss
 \textfont1=\tbmt \scriptfont1=\tbms \scriptscriptfont1=\tbmss
 \textfont2=\tbst \scriptfont2=\tbss \scriptscriptfont2=\tbsss
     \baselineskip=13dd\lineskip=0pt\lineskiplimit=0pt
     \rightskip=0pt plus4cm
     \pretolerance=10000
     \noindent
     \tbtt}
 \def \eTb{\vskip10pt
     \egroup
     \noindent
     \ignorespaces}
\newcount\section@penalty  \section@penalty=0
\newcount\subsection@penalty  \subsection@penalty=0
\newcount\subsubsection@penalty  \subsubsection@penalty=0
\def\titlea#1{\par\stepc{Tl}
    \resetcount{Tm}
    \bgroup
       \normalsize
       \bf \rightskip 0pt plus4em
       \pretolerance=20000
       \boldmath
       \setbox0=\vbox{\vskip\tabefore
          \noindent
          \arabic{Tl}.\
          \ignorespaces#1
          \vskip\taafter}
       \dimen0=\ht0\advance\dimen0 by\dp0
       \advance\dimen0 by 2\baselineskip
       \advance\dimen0 by\pagetotal
       \ifdim\dimen0>\pagegoal
          \ifdim\pagetotal>\pagegoal
          \else\eject\fi\fi
       \vskip\tabefore
       \penalty\section@penalty \global\section@penalty=-200
       \global\subsection@penalty=10007
       \noindent
       \arabic{Tl}.\
       \ignorespaces#1
       \vskip\taafter
    \egroup
    \nobreak
    \parindent=0pt
    \let\lasttitle=A%
\everypar={\parindent=\stdparindent
    \penalty\z@\let\lasttitle=N\everypar={}}%
    \ignorespaces}
\def\titleb#1{\par\stepc{Tm}
    \resetcount{Tn}
    \if N\lasttitle\else\vskip\tbbeforeback\fi
    \bgroup
       \normalsize
       \raggedright
       \pretolerance=10000
       \it
       \setbox0=\vbox{\vskip\tbbefore
          \normalsize
          \raggedright
          \pretolerance=10000
          \noindent \it \arabic{Tl}.\arabic{Tm}.\ \ignorespaces#1
          \vskip\tbafter}
       \dimen0=\ht0\advance\dimen0 by\dp0\advance\dimen0 by 2\baselineskip
       \advance\dimen0 by\pagetotal
       \ifdim\dimen0>\pagegoal
          \ifdim\pagetotal>\pagegoal
          \else \if N\lasttitle\eject\fi \fi\fi
       \vskip\tbbefore
       \if N\lasttitle \penalty\subsection@penalty \fi
       \global\subsection@penalty=-100
       \global\subsubsection@penalty=10007
       \noindent \arabic{Tl}.\arabic{Tm}.\ \ignorespaces#1
       \vskip\tbafter
    \egroup
    \nobreak
    \let\lasttitle=B%
    \parindent=0pt
    \everypar={\parindent=\stdparindent
       \penalty\z@\let\lasttitle=N\everypar={}}%
       \ignorespaces}
\def\titlec#1{\par\stepc{Tn}
    \resetcount{To}
    \if N\lasttitle\else\vskip\tcbeforeback\fi
    \bgroup
       \normalsize
       \raggedright
       \pretolerance=10000
       \setbox0=\vbox{\vskip\tcbefore
          \noindent
          \arabic{Tl}.\arabic{Tm}.\arabic{Tn}.\
          \ignorespaces#1\vskip\tcafter}
       \dimen0=\ht0\advance\dimen0 by\dp0\advance\dimen0 by 2\baselineskip
       \advance\dimen0 by\pagetotal
       \ifdim\dimen0>\pagegoal
           \ifdim\pagetotal>\pagegoal
           \else \if N\lasttitle\eject\fi \fi\fi
       \vskip\tcbefore
       \if N\lasttitle \penalty\subsubsection@penalty \fi
       \global\subsubsection@penalty=-50
       \noindent
       \arabic{Tl}.\arabic{Tm}.\arabic{Tn}.\
       \ignorespaces#1\vskip\tcafter
    \egroup
    \nobreak
    \let\lasttitle=C%
    \parindent=0pt
    \everypar={\parindent=\stdparindent
       \penalty\z@\let\lasttitle=N\everypar={}}%
       \ignorespaces}
\def\titled#1{\par\stepc{To}
    \resetcount{Tp}
    \if N\lasttitle\else\vskip\tdbeforeback\fi
    \vskip\tdbefore
    \bgroup
       \normalsize
       \if N\lasttitle \penalty-50 \fi
       \it \noindent \ignorespaces#1\unskip\
    \egroup\ignorespaces}
\def\begref#1{\par
   \unvskip
   \goodbreak\vskip\tabefore
   {\noindent\bf\ignorespaces#1%
   \par\vskip\taafter}\nobreak\let\INS=N}
\def\ref{\if N\INS\let\INS=Y\else\goodbreak\fi
   \hangindent\stdparindent\hangafter=1\noindent\ignorespaces}
% Ende der Referenzen
%

%
\def\appendix#1{\vskip\tabefore
    \vbox{\noindent{\bf Appendix #1}\vskip\taafter}%
    \global\eqnum=0\relax
    \nobreak\noindent\ignorespaces}
\let\REFEREE=N
\newbox\refereebox
\setbox\refereebox=\vbox
to0pt{\vskip0.5cm\fullline{\hrulefill\tentt\lower0.5ex
\hbox{\kern5pt referee's copy\kern5pt}\hrulefill}\vss}%
\def\refereelayout{\let\REFEREE=M\footline={\copy\refereebox}
    \message{|A referee's copy will be produced}\par
    \if N\lr\else\if R\lr \onecolumn \fi \let\lr=N \topskip=10pt\fi}

\def\utw{\smash{\rlap{\lower5pt\hbox{$\sim$}}}}
\def\udtw{\smash{\rlap{\lower6pt\hbox{$\approx$}}}}

 %reelle Zahlen
 %natuerliche Zahlen

\def\bbbc{{\mathchoice {\setbox0=\hbox{$\displaystyle\rm C$}\hbox{\hbox
to0pt{\kern0.4\wd0\vrule height0.9\ht0\hss}\box0}}
{\setbox0=\hbox{$\textstyle\rm C$}\hbox{\hbox
to0pt{\kern0.4\wd0\vrule height0.9\ht0\hss}\box0}}
{\setbox0=\hbox{$\scriptstyle\rm C$}\hbox{\hbox
to0pt{\kern0.4\wd0\vrule height0.9\ht0\hss}\box0}}
{\setbox0=\hbox{$\scriptscriptstyle\rm C$}\hbox{\hbox
to0pt{\kern0.4\wd0\vrule height0.9\ht0\hss}\box0}}}}
\def\bbbq{{\mathchoice {\setbox0=\hbox{$\displaystyle\rm Q$}\hbox{\raise
0.15\ht0\hbox to0pt{\kern0.4\wd0\vrule height0.8\ht0\hss}\box0}}
{\setbox0=\hbox{$\textstyle\rm Q$}\hbox{\raise
0.15\ht0\hbox to0pt{\kern0.4\wd0\vrule height0.8\ht0\hss}\box0}}
{\setbox0=\hbox{$\scriptstyle\rm Q$}\hbox{\raise
0.15\ht0\hbox to0pt{\kern0.4\wd0\vrule height0.7\ht0\hss}\box0}}
{\setbox0=\hbox{$\scriptscriptstyle\rm Q$}\hbox{\raise
0.15\ht0\hbox to0pt{\kern0.4\wd0\vrule height0.7\ht0\hss}\box0}}}}
\def\bbbt{{\mathchoice {\setbox0=\hbox{$\displaystyle\rm
T$}\hbox{\hbox to0pt{\kern0.3\wd0\vrule height0.9\ht0\hss}\box0}}
{\setbox0=\hbox{$\textstyle\rm T$}\hbox{\hbox
to0pt{\kern0.3\wd0\vrule height0.9\ht0\hss}\box0}}
{\setbox0=\hbox{$\scriptstyle\rm T$}\hbox{\hbox
to0pt{\kern0.3\wd0\vrule height0.9\ht0\hss}\box0}}
{\setbox0=\hbox{$\scriptscriptstyle\rm T$}\hbox{\hbox
to0pt{\kern0.3\wd0\vrule height0.9\ht0\hss}\box0}}}}
\def\bbbs{{\mathchoice
{\setbox0=\hbox{$\displaystyle     \rm S$}\hbox{\raise0.5\ht0\hbox
to0pt{\kern0.35\wd0\vrule height0.45\ht0\hss}\hbox
to0pt{\kern0.55\wd0\vrule height0.5\ht0\hss}\box0}}
{\setbox0=\hbox{$\textstyle        \rm S$}\hbox{\raise0.5\ht0\hbox
to0pt{\kern0.35\wd0\vrule height0.45\ht0\hss}\hbox
to0pt{\kern0.55\wd0\vrule height0.5\ht0\hss}\box0}}
{\setbox0=\hbox{$\scriptstyle      \rm S$}\hbox{\raise0.5\ht0\hbox
to0pt{\kern0.35\wd0\vrule height0.45\ht0\hss}\raise0.05\ht0\hbox
to0pt{\kern0.5\wd0\vrule height0.45\ht0\hss}\box0}}
{\setbox0=\hbox{$\scriptscriptstyle\rm S$}\hbox{\raise0.5\ht0\hbox
to0pt{\kern0.4\wd0\vrule height0.45\ht0\hss}\raise0.05\ht0\hbox
to0pt{\kern0.55\wd0\vrule height0.45\ht0\hss}\box0}}}}
\def\bbbz{{\mathchoice {\hbox{$\sf\textstyle Z\kern-0.4em Z$}}
{\hbox{$\sf\textstyle Z\kern-0.4em Z$}}
{\hbox{$\sf\scriptstyle Z\kern-0.3em Z$}}
{\hbox{$\sf\scriptscriptstyle Z\kern-0.2em Z$}}}}
\def\diameter{{\ifmmode\oslash\else$\oslash$\fi}}

\def\vec#1{{\boldmath
\textfont0=\bfIXf\scriptfont0=\bfVIf\scriptscriptfont0=\bfVf
\ifmmode
\mathchoice{\hbox{$\displaystyle#1$}}{\hbox{$\textstyle#1$}}
{\hbox{$\scriptstyle#1$}}{\hbox{$\scriptscriptstyle#1$}}\else
$#1$\fi}}
\def\tens#1{\ifmmode
\mathchoice{\hbox{$\displaystyle\sf#1$}}{\hbox{$\textstyle\sf#1$}}
{\hbox{$\scriptstyle\sf#1$}}{\hbox{$\scriptscriptstyle\sf#1$}}\else
$\sf#1$\fi}
\newcount\sterne \sterne=0
\newdimen\fullhead
{\catcode`@=11    % use @ as a normal character
\def\newtoks{\alloc@5\toks\toksdef\@cclvi}
\outer\gdef\makenewtoks#1{\newtoks#1#1={ ????? }}}
\makenewtoks\DATE
\makenewtoks\MAINTITLE
\makenewtoks\SUBTITLE
\makenewtoks\AUTHOR
\makenewtoks\INSTITUTE
\makenewtoks\ABSTRACT
\makenewtoks\KEYWORDS
\makenewtoks\THESAURUS
\makenewtoks\OFFPRINTS
\newlinechar=`\| %
\let\INS=N%
{\catcode`\@=\active
\gdef@#1{\if N\INS $^{#1}$\else\if
E\INS\hangindent0.5\stdparindent\hangafter=1%
\noindent\hbox to0.5\stdparindent{$^{#1}$\hfil}\let\INS=Y\ignorespaces
\else\par\hangindent0.5\stdparindent\hangafter=1
\noindent\hbox to0.5\stdparindent{$^{#1}$\hfil}\ignorespaces\fi\fi}%
}%
\def\mehrsterne{\global\advance\sterne by1\relax}%
\def\footnoterule{\kern-3pt\hrule width 2true cm\kern2.6pt}% Trennlinie
\def\makeOFFPRINTS#1{\bgroup\normalsize
       \hsize=19.5cc
       \baselineskip=10dd\lineskiplimit=0pt\lineskip=0pt
       \def\textindent##1{\noindent{\it Send offprint
          requests to\/}: }\relax
       \vfootnote{nix}{\ignorespaces#1}\egroup}
\def\makesterne{\count254=0\loop\ifnum\count254<\sterne
\advance\count254 by1\star\repeat}
\def\FOOTNOTE#1{\bgroup
       \ifhmode\unskip\fi
       \mehrsterne$^{\makesterne}$\relax
       \normalsize
       \hsize=19.5cc
       \baselineskip=10dd\lineskiplimit=0pt\lineskip=0pt
       \def\textindent##1{\noindent\hbox
       to\stdparindent{##1\hss}}\relax
       \vfootnote{$^{\makesterne}$}{\ignorespaces#1}\egroup}
\def\fonote#1{\ifhmode\unskip\fi
       \mehrsterne$^{\the\sterne}$\bgroup
       \normalsize
       \hsize=19.5cc
       \def\textindent##1{\noindent\hbox
       to\stdparindent{##1\hss}}\relax
       \vfootnote{$^{\the\sterne}$}{\ignorespaces#1}\egroup}
\def\missmsg#1{\message{|Missing #1 }}
\def\tstmiss#1#2#3#4#5{%
\edef\test{\the #1}%
\ifx\test\missing%
  #2\relax%  message
  #3%   action if missing
\else
  \ifx\test\missingi%
    #2\relax%  message
    #3%   action if missing
  \else #4%  action if existing
  \fi
\fi
#5%   action at any rate
}%
\def\maketitle{\paglay%
\def\missing{ ????? }%
\def\missingi{ }%
{\parskip=0pt\relax
\setbox0=\vbox{\hsize=\fullhsize\null\vskip2truecm
\tstmiss%
  {\MAINTITLE}%
  {}%
  {\global\MAINTITLE={MAINTITLE should be given}}%
  {}%
  {%   write MAINTITLE:
   \aTa\ignorespaces\the\MAINTITLE\eTa}%
\tstmiss%
  {\SUBTITLE}%
  {}%
  {}%
  {%   write SUBTITLE:
   \aTb\ignorespaces\the\SUBTITLE\eTb}%
  {}%
\tstmiss%
  {\AUTHOR}%
  {}%
  {\AUTHOR={Name(s) and initial(s) of author(s) should be given}}
  {}%
  {%   write AUTHOR:
\noindent{\bf\ignorespaces\the\AUTHOR\vskip4pt}}%
\tstmiss%
  {\INSTITUTE}%
  {}%
  {\INSTITUTE={Address(es) of author(s) should be given.}}%
  {}%
  {%   write INSTITUTE:
   \let\INS=E
\noindent\ignorespaces\the\INSTITUTE\vskip10pt}%
\tstmiss%
  {\DATE}%
  {}%
  {\DATE={$[$the date of receipt and acceptance should be inserted
later$]$}}%
  {}%
  {%   write DATE:
{\noindent\ignorespaces\the\DATE\vskip21pt}\bf A}%
}%
\global\fullhead=\ht0\global\advance\fullhead by\dp0
\global\advance\fullhead by10pt\global\sterne=0
{\hsize=19.5cc\null\vskip2truecm
\tstmiss%
  {\OFFPRINTS}%
  {}%
  {}%
  {\makeOFFPRINTS{\the\OFFPRINTS}}%
  {}%
\hsize=\fullhsize
\tstmiss%
  {\MAINTITLE}%
  {\missmsg{MAINTITLE}}%
  {\global\MAINTITLE={MAINTITLE should be given}}%
  {}%
  {%   write MAINTITLE:
   \aTa\ignorespaces\the\MAINTITLE\eTa}%
\tstmiss%
  {\SUBTITLE}%
  {}%
  {}%
  {%   write SUBTITLE:
   \aTb\ignorespaces\the\SUBTITLE\eTb}%
  {}%
\tstmiss%
  {\AUTHOR}%
  {\missmsg{name(s) and initial(s) of author(s)}}%
  {\AUTHOR={Name(s) and initial(s) of author(s) should be given}}
  {}%
  {%   write AUTHOR:
\noindent{\bf\ignorespaces\the\AUTHOR\vskip4pt}}%
\tstmiss%
  {\INSTITUTE}%
  {\missmsg{address(es) of author(s)}}%
  {\INSTITUTE={Address(es) of author(s) should be given.}}%
  {}%
  {%   write INSTITUTE:
   \let\INS=E
\noindent\ignorespaces\the\INSTITUTE\vskip10pt}%
\catcode`\@=12
\tstmiss%
  {\DATE}%
  {\message{|The date of receipt and acceptance should be inserted
later.}}%
  {\DATE={$[$the date of receipt and acceptance should be inserted
later$]$}}%
  {}%
  {%   write DATE:
{\noindent\ignorespaces\the\DATE\vskip21pt}}%
}%
\tstmiss%
  {\THESAURUS}%
  {\message{|Thesaurus codes are not given.}}%
  {\global\THESAURUS={missing; you have not inserted them}}%
  {}%
  {}%
\if M\REFEREE\let\REFEREE=Y
\normalbaselineskip=2\normalbaselineskip
\normallineskip=2\normallineskip\normalbaselines\fi
\tstmiss%
  {\ABSTRACT}%
  {\missmsg{ABSTRACT}}%
  {\ABSTRACT={Not yet given.}}%
  {}%
  {\noindent{\bf Abstract. }\ignorespaces\the\ABSTRACT\vskip0.5true cm}%
\def\strich{\par
\vbox to0pt{\hrule width\hsize\vss}\vskip-1.2\baselineskip
\vskip0pt plus3\baselineskip\relax}%
\tstmiss%
  {\KEYWORDS}%
  {\missmsg{KEYWORDS}}%
  {\KEYWORDS={Not yet given.}}%
  {}%
  {\noindent{\bf Key words: }\ignorespaces\the\KEYWORDS
  \strich}%
\global\sterne=0
}}%Ende von maketitle
\newdimen\@txtwd  \@txtwd=\hsize
\newdimen\@txtht  \@txtht=\vsize
\newdimen\@colht  \@colht=\vsize
\newdimen\@colwd  \@colwd=-1pt
\newdimen\@colsavwd
\newcount\in@t \in@t=0
\def\initlr{\if N\lr \ifdim\@colwd<0pt \global\@colwd=\hsize \fi
   \else\global\let\lr=L\ifdim\@colwd<0pt \global\@colwd=\hsize
      \global\divide\@colwd\tw@ \global\advance\@colwd by -10pt
   \fi\fi\global\advance\in@t by 1}
\def\setuplr#1#2#3{\let\lr=O \ifx#1\lr\global\let\lr=N
      \else\global\let\lr=L\fi
   \@txtht=\vsize \@colht=\vsize \@txtwd=#2 \@colwd=#3
   \if N\lr \else\multiply\@colwd\tw@ \fi
   \ifdim\@colwd>\@txtwd\if N\lr
        \errmessage{The text width is less than the column width}%
      \else
        \errmessage{The text width is less the two times the column width}%
      \fi \global\@colwd=\@txtwd
      \if N\lr\divide\@colwd by 2\fi
   \else \global\@colwd=#3 \fi \initlr \@colsavwd=#3
   \global\@insmx=\@txtht
   \global\hsize=\@colwd}
\def\twocolumns{\@fillpage\eject\global\let\lr=L \@makecolht
   \global\@colwd=\@colsavwd \global\hsize=\@colwd}
\def\onecolumn{\@fillpage\eject\global\let\lr=N \@makecolht
   \global\@colwd=\@txtwd \global\hsize=\@colwd}
\def\newpage{\@fillpage\eject}
\def\@fillpage{\vfill\supereject\if R\lr \null\vfill\eject\fi}

\newbox\@leftcolumn
\newbox\@rightcolumn
\newbox\@outputbox
\newbox\@tempboxa
\newbox\@keepboxa
\newbox\@keepboxb
\newbox\@bothcolumns
\newbox\@savetopins
\newbox\@savetopright
\newcount\verybad \verybad=1010
\def\@makecolumn{\ifnum \in@t<1\initlr\fi
   \ifnum\outputpenalty=\the\verybad1  %%% i.e. 10101 if \verybad=1010
      \if L\lr\else\advance\pageno by1\fi
      \message{Warning: There is a 'widow' line
      at the top of page \the\pageno\if R\lr (left)\fi.
      This is unacceptable.} \if L\lr\else\advance\pageno by-1\fi \fi
   \ifnum\outputpenalty=\the\verybad2
      \message{Warning: There is a 'club' line
      at the bottom of page \the\pageno\if L\lr(left)\fi.
      This is unacceptable.} \fi
   \if L\lr \ifvoid\@savetopins\else\@colht=\@txtht\fi \fi
   \if R\lr \ifvoid\@bothcolumns \ifvoid\@savetopright
       \else\@colht=\@txtht\fi\fi\fi
   \global\setbox\@outputbox
   \vbox to\@colht{\boxmaxdepth\maxdepth
   \if L\lr \ifvoid\@savetopins\else\unvbox\@savetopins\fi \fi
   \if R\lr \ifvoid\@bothcolumns \ifvoid\@savetopright\else
       \unvbox\@savetopright\fi\fi\fi
   \ifvoid\topins\else\ifnum\count\topins>0
         \ifdim\ht\topins>\@colht
            \message{|Error: Too many or too large single column
            box(es) on this page.}\fi
         \unvbox\topins
      \else
         \global\setbox\@savetopins=\vbox{\ifvoid\@savetopins\else
         \unvbox\@savetopins\penalty-500\fi \unvbox\topins} \fi\fi
   \dimen@=\dp\@cclv \unvbox\@cclv % open up \box255
   \ifvoid\bottomins\else\unvbox\bottomins\fi
   \ifvoid\footins\else % footnote info is present
     \vskip\skip\footins
     \footnoterule
     \unvbox\footins\fi
   \ifr@ggedbottom \kern-\dimen@ \vfil \fi}%
}
\def\@outputpage{\@dooutput{\lr}}
\def\@colbox#1{\hbox to\@colwd{\box#1\hss}}
\def\@dooutput#1{\global\topskip=10pt
  \ifdim\ht\@bothcolumns>\@txtht
    \if #1N
       \unvbox\@outputbox
    \else
       \unvbox\@leftcolumn\unvbox\@outputbox
    \fi
    \global\setbox\@tempboxa\vbox{\hsize=\@txtwd\makeheadline
       \vsplit\@bothcolumns to\@txtht
       \makefootline\hsize=\@colwd}%
    \message{|Error: Too many double column boxes on this page.}%
    \shipout\box\@tempboxa\advancepageno
    \unvbox255 \penalty\outputpenalty
  \else
    \global\setbox\@tempboxa\vbox{\hsize=\@txtwd\makeheadline
       \ifvoid\@bothcolumns\else\unvbox\@bothcolumns\fi
       \hsize=\@colwd
       \if #1N
          \hbox to\@txtwd{\@colbox{\@outputbox}\hfil}%
       \else
          \hbox to\@txtwd{\@colbox{\@leftcolumn}\hfil\@colbox{\@outputbox}}%
       \fi
       \hsize=\@txtwd\makefootline\hsize=\@colwd}%
    \shipout\box\@tempboxa\advancepageno
  \fi
  \ifnum \special@pages>0 \s@count=100 \page@command
      \xdef\page@command{}\global\special@pages=0 \fi
  }
\def\balance@right@left{\dimen@=\ht\@leftcolumn
    \advance\dimen@ by\ht\@outputbox
    \advance\dimen@ by\ht\springer@macro
    \dimen2=\z@ \global\the@end=0
    \ifdim\dimen@>70pt\setbox\z@=\vbox{\unvbox\@leftcolumn
          \unvbox\@outputbox}%
       \loop
          \dimen@=\ht\z@
          \advance\dimen@ by0.5\topskip
          \advance\dimen@ by\baselineskip
          \advance\dimen@ by\ht\springer@macro
          \advance\dimen@ by\dimen2
          \divide\dimen@ by2
          \splittopskip=\topskip
          % Now split it to two parts of about the same height
          {\vbadness=10000
             \global\setbox3=\copy\z@
             \global\setbox1=\vsplit3 to \dimen@}%
          \dimen1=\ht3 \advance\dimen1 by\ht\springer@macro
       \ifdim\dimen1>\ht1 \advance\dimen2 by\baselineskip\repeat
       \dimen@=\ht1
       % Restore the column boxes and adjust
       \global\setbox\@leftcolumn
          \hbox to\@colwd{\vbox to\@colht{\vbox to\dimen@{\unvbox1}\vfil}}%
       \global\setbox\@outputbox
          \hbox to\@colwd{\vbox to\@colht{\vbox to\dimen@{\unvbox3
             \vfill\box\springer@macro}\vfil}}%
    \else
       \setbox\@leftcolumn=\vbox{unvbox\@leftcolumn\bigskip
          \box\springer@macro}%
    \fi}
\newinsert\bothins
\newbox\rightins
\skip\bothins=\z@skip
\count\bothins=1000
\dimen\bothins=\@txtht \advance\dimen\bothins by -\bigskipamount
\def\bothtopinsert{\par\begingroup\setbox\z@\vbox\bgroup
    \hsize=\@txtwd\parskip=0pt\par\noindent\bgroup}
\def\endbothinsert{\egroup\egroup
  \if R\lr
    \right@nsert
  \else    % L\lr or N\lr
    \dimen@=\ht\z@ \advance\dimen@ by\dp\z@ \advance\dimen@ by\pagetotal
    \advance\dimen@ by \bigskipamount \advance\dimen@ by \topskip
    \advance\dimen@ by\ht\topins \advance\dimen@ by\dp\topins
    \advance\dimen@ by\ht\bottomins \advance\dimen@ by\dp\bottomins
    \advance\dimen@ by\ht\@savetopins \advance\dimen@ by\dp\@savetopins
    \ifdim\dimen@>\@colht\right@nsert\else\left@nsert\fi
  \fi  \endgroup}
\def\right@nsert{\global\setbox\rightins\vbox{\ifvoid\rightins
    \else\unvbox\rightins\fi\penalty100
    \splittopskip=\topskip
    \splitmaxdepth\maxdimen \floatingpenalty200
    \dimen@\ht\z@ \advance\dimen@\dp\z@
    \box\z@\nobreak\bigskip}}
\def\left@nsert{\insert\bothins{\penalty100
    \splittopskip=\topskip
    \splitmaxdepth\maxdimen \floatingpenalty200
    \box\z@\nobreak\bigskip}
    \@makecolht}
\newdimen\@insht    \@insht=\z@
\newdimen\@insmx    \@insmx=\vsize
\def\@makecolht{\global\@colht=\@txtht \@compinsht
    \global\advance\@colht by -\@insht \global\vsize=\@colht
    \global\dimen\topins=\@colht}
\def\@compinsht{\if R\lr
       \dimen@=\ht\@bothcolumns \advance\dimen@ by\dp\@bothcolumns
       \ifvoid\@bothcolumns \advance\dimen@ by\ht\@savetopright
          \advance\dimen@ by\dp\@savetopright \fi
    \else
       \dimen@=\ht\bothins \advance\dimen@ by\dp\bothins
       \advance\dimen@ by\ht\@savetopins \advance\dimen@ by\dp\@savetopins
    \fi
    \ifdim\dimen@>\@insmx
       \global\@insht=\dimen@
    \else\global\@insht=\dimen@
    \fi}
\newinsert\bottomins
\skip\bottomins=\z@skip
\count\bottomins=1000
\xdef\page@command{}
\newcount\s@count
\newcount\special@pages \special@pages=0
\def\specialpage#1{\global\advance\special@pages by1
    \global\s@count=\special@pages
    \global\advance\s@count by 100
    \global\setbox\s@count
    \vbox to\@txtht{\hsize=\@txtwd\parskip=0pt
    \par\noindent\noexpand#1\vfil}%
    \def\protect{\noexpand\protect\noexpand}%
    \xdef\page@command{\page@command
         \protect\global\advance\s@count by1
         \protect\begingroup
         \protect\setbox\z@\vbox{\protect\makeheadline
                                    \protect\box\s@count
            \protect\makefootline}%
         \protect{\shipout\box\z@}%
         \protect\endgroup\protect\advancepageno}%
    \let\protect=\relax
   }
\def\@startins{\vskip \topskip\hrule height\z@
   \nobreak\vskip -\topskip\vskip3.7pt}
\let\retry=N
\output={\@makecolht \global\topskip=10pt \let\retry=N%
   \ifnum\count\topins>0 \ifdim\ht\topins>\@colht
       \global\count\topins=0 \global\let\retry=Y%
       \unvbox\@cclv \penalty\outputpenalty \fi\fi
   \if N\retry
    \if N\lr     % this is for single column output
       \@makecolumn
       \ifnum\the@end>0
          \setbox\z@=\vbox{\unvcopy\@outputbox}%
          \dimen@=\ht\z@ \advance\dimen@ by\ht\springer@macro
          \ifdim\dimen@<\@colht
             \setbox\@outputbox=\vbox to\@colht{\box\z@
             \unskip\vskip12pt plus0pt minus12pt
             \box\springer@macro\vfil}%
          \else \box\springer@macro \fi
          \global\the@end=0
       \fi
       \ifvoid\bothins\else\global\setbox\@bothcolumns\box\bothins\fi
       \@outputpage
       \ifvoid\rightins\else
       %  Hold \rightins back if there is already a \@savetopins
       \ifvoid\@savetopins\insert\bothins{\unvbox\rightins}\fi
       \fi
    \else
       \if L\lr    % this is the left of two columns
          \@makecolumn
          \global\setbox\@leftcolumn\box\@outputbox \global\let\lr=R%
          \ifnum\pageno=1
             \message{|[left\the\pageno]}%
          \else
             \message{[left\the\pageno]}\fi
          \ifvoid\bothins\else\global\setbox\@bothcolumns\box\bothins\fi
          \global\dimen\bothins=\z@
          \global\count\bothins=0
          \ifnum\pageno=1
             \global\topskip=\fullhead\fi
       \else    % the right column
          \@makecolumn
          \ifnum\the@end>0\ifnum\pageno>1\balance@right@left\fi\fi
          \@outputpage \global\let\lr=L%
          \global\dimen\bothins=\maxdimen
          \global\count\bothins=1000
          \ifvoid\rightins\else
          %  Hold \rightins back if there is already a \@savetopins
             \ifvoid\@savetopins \insert\bothins{\unvbox\rightins}\fi
          \fi
       \fi
    \fi
    \global\let\last@insert=N \put@default
    \ifnum\outputpenalty>-\@MM\else\dosupereject\fi
    \ifvoid\@savetopins\else
      \ifdim\ht\@savetopins>\@txtht
        \global\setbox\@tempboxa=\box\@savetopins
        \global\setbox\@savetopins=\vsplit\@tempboxa to\@txtht
        \global\setbox\@savetopins=\vbox{\unvbox\@savetopins}%
        \global\setbox\@savetopright=\box\@tempboxa \fi
    \fi
    \@makecolht
    \global\count\topins=1000
   \fi
   }
\if N\lr
   \setuplr{O}{\fullhsize}{\hsize}% O = one column
\else
   \setuplr{T}{\fullhsize}{\hsize}% T = two columns
\fi
\def\put@default{\global\let\insert@here=Y
   \global\let\insert@at@the@bottom=N}%
\def\puthere{\global\let\insert@here=Y%
    \global\let\insert@at@the@bottom=N}
\def\putattop{\global\let\insert@here=N%
    \global\let\insert@at@the@bottom=N}
\def\putatbottom{\global\let\insert@here=N%
    \global\let\insert@at@the@bottom=X}
\put@default
\let\last@insert=N
\def\end@skip{\smallskip}
\newdimen\min@top
\newdimen\min@here
\newdimen\min@bot
\min@top=10cm
\min@here=4cm
\min@bot=\topskip
\def\figfuzz{\vskip 0pt plus 6pt minus 3pt}  % more flexible spacing
%--------------------------------------------------------------------
\def\check@here@and@bottom#1{\relax
   \ifvoid\topins\else       \global\let\insert@here=N\fi
   \if B\last@insert         \global\let\insert@here=N\fi
   \if T\last@insert         \global\let\insert@here=N\fi
   \ifdim #1<\min@bot        \global\let\insert@here=N\fi
   \ifdim\pagetotal>\@colht  \global\let\insert@here=N\fi
   \ifdim\pagetotal<\min@here\global\let\insert@here=N\fi
   \if X\insert@at@the@bottom\global\let\insert@at@the@bottom=Y
     \else\if T\last@insert  \global\let\insert@at@the@bottom=N\fi
          \if H\last@insert  \global\let\insert@at@the@bottom=N\fi
          \ifvoid\topins\else\global\let\insert@at@the@bottom=N\fi\fi
   \ifdim #1<\min@bot        \global\let\insert@at@the@bottom=N\fi
   \ifdim\pagetotal>\@colht  \global\let\insert@at@the@bottom=N\fi
   \ifdim\pagetotal<\min@top \global\let\insert@at@the@bottom=N\fi
   \ifvoid\bottomins\else    \global\let\insert@at@the@bottom=Y\fi
   \if Y\insert@at@the@bottom\global\let\insert@here=N\fi }
\def\single@column@insert#1{\relax
   \setbox\@tempboxa=\vbox{#1}%
   \dimen@=\@colht \advance\dimen@ by -\pagetotal
   \advance\dimen@ by-\ht\@tempboxa \advance\dimen0 by-\dp\@tempboxa
   \advance\dimen@ by-\ht\topins \advance\dimen0 by-\dp\topins
   \check@here@and@bottom{\dimen@}%
   \if Y\insert@here
      \par  % The insertion forces a new paragraph in this case.
      \midinsert\figfuzz\relax     %%%%%%%%%\bigskip
      \box\@tempboxa\end@skip\figfuzz\endinsert
      \global\let\last@insert=H
   \else \if Y\insert@at@the@bottom
      \begingroup\insert\bottomins\bgroup\if B\last@insert\end@skip\fi
      \floatingpenalty=20000\figfuzz\bigskip\box\@tempboxa\egroup\endgroup
      \global\let\last@insert=B
   \else
      \topinsert\box\@tempboxa\end@skip\figfuzz\endinsert
      \global\let\last@insert=T
   \fi\fi\put@default\ignorespaces}
\def\begfig#1cm#2\endfig{\single@column@insert{\@startins\rahmen{#1}#2}%
\ignorespaces}
\def\begfigwid#1cm#2\endfig{\relax
   \if N\lr  % Here the only difference to \begfig is the larger \hsize
      {\hsize=\fullhsize \begfig#1cm#2\endfig}%
   \else
      \setbox0=\vbox{\hsize=\fullhsize\bigskip#2\smallskip}%
      \dimen0=\ht0\advance\dimen0 by\dp0
      \advance\dimen0 by#1cm
      \advance\dimen0by7\normalbaselineskip\relax
      \ifdim\dimen0>\@txtht
         \message{|Figure plus legend too high, will try to put it on a
                  separate page. }%
         \begfigpage#1cm#2\endfig
      \else
         \bothtopinsert\line{\vbox{\hsize=\fullhsize
         \@startins\rahmen{#1}#2\smallskip}\hss}\figfuzz\endbothinsert
      \fi
   \fi}
\def\begfigside#1cm#2cm#3\endfig{\relax
   \if N\lr  % Here the only difference to \begfig is the larger \hsize
      {\hsize=\fullhsize \begfig#1cm#3\endfig}%
   \else
      \dimen0=#2true cm\relax
      \ifdim\dimen0<\hsize
         \message{|Your figure fits in a single column; why don't|you use
                  \string\begfig\space instead of \string\begfigside? }%
      \fi
      \dimen0=\fullhsize
      \advance\dimen0 by-#2true cm
      \advance\dimen0 by-1true cc\relax
      \bgroup
         \ifdim\dimen0<8true cc\relax
            \message{|No sufficient room for the legend;
                     using \string\begfigwid. }%
            \begfigwid #1cm#3\endfig
         \else
            \ifdim\dimen0<10true cc\relax
               \message{|Room for legend to narrow;
                        legend will be set raggedright. }%
               \rightskip=0pt plus 2cm\relax
            \fi
            \setbox0=\vbox{\def\figure##1##2{\vbox{\hsize=\dimen0\relax
                           \@startins\noindent\petit{\bf
                           Fig.\ts##1\unskip.\ }\ignorespaces##2\par}}%
                           #3\unskip}%
            \ifdim#1true cm<\ht0\relax
               \message{|Text of legend higher than figure; using
                        \string\begfig. }%
               \begfigwid #1cm#3\endfig
            \else
               \def\figure##1##2{\vbox{\hsize=\dimen0\relax
                                       \@startins\noindent\petit{\bf
                                       Fig.\ts##1\unskip.\
                                       }\ignorespaces##2\par}}%
               \bothtopinsert\line{\vbox{\hsize=#2true cm\relax
               \@startins\rahmen{#1}}\hss#3\unskip}\figfuzz\endbothinsert
            \fi
         \fi
      \egroup
   \fi\ignorespaces}
\def\begfigpage#1cm#2\endfig{\specialpage{\@startins
   \vskip3.7pt\rahmen{#1}#2}\ignorespaces}%
\def\begtab#1cm#2\endtab{\single@column@insert{#2\rahmen{#1}}\ignorespaces}
\let\begtabempty=\begtab
\def\begtabfull#1\endtab{\single@column@insert{#1}\ignorespaces}
\def\begtabemptywid#1cm#2\endtab{\relax
   \if N\lr
      {\hsize=\fullhsize \begtabempty#1cm#2\endtab}%
   \else
      \bothtopinsert\line{\vbox{\hsize=\fullhsize
      #2\rahmen{#1}}\hss}\medskip\endbothinsert
   \fi\ignorespaces}
\def\begtabfullwid#1\endtab{\relax
   \if N\lr
      {\hsize=\fullhsize \begtabfull#1\endtab}%
   \else
      \bothtopinsert\line{\vbox{\hsize=\fullhsize
      \noindent#1}\hss}\medskip\endbothinsert
   \fi\ignorespaces}
\def\begtabpage#1\endtab{\specialpage{#1}\ignorespaces}
\catcode`\@=\active   % This is reset by the \maketitle macro

\input psfig.tex
\voffset=1.0truecm
%\refereelayout
\MAINTITLE{Leptonic origin of TeV gamma-rays from Supernova Remnants}
\AUTHOR{Martin Pohl}
\INSTITUTE{Max-Planck-Institut f\"ur extraterrestrische
Physik, Postfach 1603, 85740 Garching, Germany}
\ABSTRACT{The lineless power-law emission observed by ASCA
from the northeastern rim of
the supernova remnant SN1006 has recently been interpreted as synchrotron
radiation of electrons with energies around 100 TeV. In this letter
we calculate the flux of inverse Compton emission at TeV photon energies that
is a natural consequence of the existence of such high energy electrons
and the cosmic microwave background. We find that the predicted flux is
near the present sensitivity limit of the southern \v Cerenkov telescope 
CANGAROO, and should be detectable with the next performance improvements.
The spectrum of SN1006 at a few TeV will be very soft.

The existence of such highest energy electrons in SN1006 may not be a unique
to this remnant.
We can therefore conclude that the detection of TeV $\gamma$-ray
emission in any supernova remnant does not necessarily provide evidence
for a large number of cosmic ray nucleons in these objects, and thus is no 
simple test of cosmic ray origin as far as nucleons are concerned.}
\THESAURUS{02.01.1; 09.03.2; 09.19.2; 13.07.3}
\OFFPRINTS{\hphantom{http://www.gamma.mpe-garching} http://www.gamma.mpe-garching.mpg.de/$\sim$mkp/mkp.html}
\DATE{received: ?; accepted: ? }
\maketitle
%**************************************************************************
\titlea{Introduction}

In a recent paper Koyama et al. (1995) claimed evidence for shock
accelerated electrons with energies around 100 TeV in the supernova remnant 
SN1006. ASCA X-ray observations of the northeastern rim show a lineless
X-ray continuum with a power-law spectrum up to 8 keV, in contrast to the
spectrum of the interior of the remnant which is dominated by emission lines
characteristic of highly ionized material. This power-law emission is 
interpreted as synchrotron radiation of high-energy electrons in the 
ambient magnetic fields. For a field strength of 10$\,\mu $G the required
electron energy would be of order 100 TeV. The synchrotron origin is
supported by a correlation between X-ray and radio surface brightness
along the northeastern rim. Adding the southwestern counterpart the
rims provide 75\% of the integrated flux above 1 keV.

The energy spectral index of the power-law emission is $\alpha =1.95\pm 0.2$,
similar to the integrated spectrum seen by GINGA (Ozaki et al. 1994), but steeper than the $\alpha \simeq 1.2$ from earliest observations
(Becker et al. 1980). The old spectral index value was very attractive as it
allowed modelling of the radio-to-X-ray spectrum as synchrotron emission of
electrons which are accelerated in a shock wave against losses due to
adiabatic expansion and synchrotron radiation (Reynolds and Chevalier 1981;
Ammosov et al. 1994). An energy spectral index around 2 at X-ray energies 
will require more complicated models, for example a cut-off in the
electron spectrum which is smoothed out in the synchrotron emission
by variations of the magnetic field strength.

One may be attracted to attribute the observed X-ray spectrum to
inverse Compton emission. Given the obvious spectral differences
between the radio and X-ray emission and the lack of target photons
with energies of less than a few $10^{-5}\,$eV the only reasonable scenario
would involve scattering of NIR and optical photons by electrons with 
Lorentz factors between 10 and 100. To match the observed luminosities at
X-rays and radio frequencies the electron spectrum would have to be extremely
soft below a critical Lorentz factor of $\gamma_c=1000$. If not invoking
a new population of electrons the only reasonable model
for this would be acceleration against strong Coulomb and
ionization losses which leads to an additional factor
exp($\gamma_c / \gamma$) in the electron number density and a
soft synchrotron spectrum with power-law character. However, Coulomb
and ionization losses at $\gamma_c=1000$ on a time scale of much less
than the age of SN1006 imply a density of $n\gg 3\cdot 10^4\ {\rm cm^{-3}}$
which appears excluded by H$I$ data and the lack of free-free absorption. 

In this letter we propose a test for the hypothesis that electrons are
frequently accelerated to energies around 100 TeV in 
the shells of supernova remnants. Electrons of this energy will
upscatter photons of the microwave background to energies of a few TeV,
well in the energy range of \v Cerenkov telescopes. The flux in $\gamma$-rays
of energy around 10 TeV
is independent of the spectral index since we see particles of similar
energy in X-rays and TeV $\gamma$-rays. The only unknown variable is
the magnetic field strength which controls the synchrotron emissivity.
In the next section we will calculate the flux of TeV $\gamma$-rays
for two values of the X-ray spectral index to estimate also the effect
of a slow turn-over in the electron spectrum. The result will be discussed 
in comparison to the sensitivity of the \v Cerenkov telescope CANGAROO
in section 3.

\titlea{Emission from high-energy electrons}
\titleb{Synchrotron radiation at X-rays}

We have estimated the integral X-ray emission from SN1006 from the results
of GINGA (Ozaki et al. 1994) and TENMA (Koyama et al. 1987). Attributing
40\% of this to the northeastern rim we obtain for its
flux
$$I_x = 10^{-3} \left({{E}\over {3\ {\rm keV}}}\right)^{-3}\ {\rm
ph./cm^2/sec/keV}\ \eqno(1)$$
We have performed the calculations also for a photon spectral index
of 2.5 to account for the spectral
uncertainties and possible spectral variations.
With
$$N = C \gamma^{-r} = C (m c^2)^{(r-1)} E^{-r}\ \eqno(2)$$
as the total number spectrum
of relativistic electrons we can calculate the synchrotron flux at X-ray
energies for a randomly oriented magnetic field (Rybicki and Lightman 1979)
\goodbreak
$$I(E)\simeq {{\sqrt{3} e^2 C h^{s-2}}\over {12\,s D^2 c}}
\Gamma \left({{3s+8}\over {6}}\right)
\Gamma \left({{3s-2}\over {6}}\right)
\left({{3eB}\over {8mc\,E}}\right)^{s}$$ $$\qquad\qquad\qquad\qquad
\qquad\qquad {\rm ph./cm^2/sec/erg}\ \eqno(3)$$
where $s=(r+1)/2$ is the photon spectral index, $D$ is the distance, and
$h$ is Planck's constant. To match the observed photon flux the number of 
electrons is required to be
$${{C}\over {D^2}} \simeq \cases{ 6\cdot 10^{31}\  B_{-5}^{-3}\qquad&
for\quad s=3\cr\cr
5\cdot 10^{23}\  B_{-5}^{-2.5}\ \ &
for\quad s=2.5\cr}\ \eqno(4)$$
where $B_{-5}$ is the magnetic field strength in units of $10\ \mu G$.

\titleb{Inverse Compton emission at TeV energies}

Due to the Klein-Nishina cut-off only scattering of microwave background
photons is important. The differential cross section for the up-scattering
of a photon with incident energy $\epsilon$ to energy $E_\gamma$ by 
elastic collision with an electron of energy $E$ is given by
(Blumenthal and Gould 1970)
$$\sigma(E_\gamma, \epsilon, E) = {{3\, \sigma_T (m c^2)^2}\over
{4\, \epsilon E^2}}$$ $$\qquad \times \ \left[ 2q {\rm ln}q +(1+2q)(1-q)
+0.5{{(\Gamma_e q)^2 (1-q)}\over {1+\Gamma_e q}}\right] \ \eqno(5)$$
where
$$q={E_\gamma \over {\Gamma_e (E-E_\gamma)}}\quad {\rm and}\quad
\Gamma_e = {{4\, \epsilon E}\over {m^2 c^4}}$$
\vskip0.4truecm
%\begfig 8.5 cm
{\psfig{figure=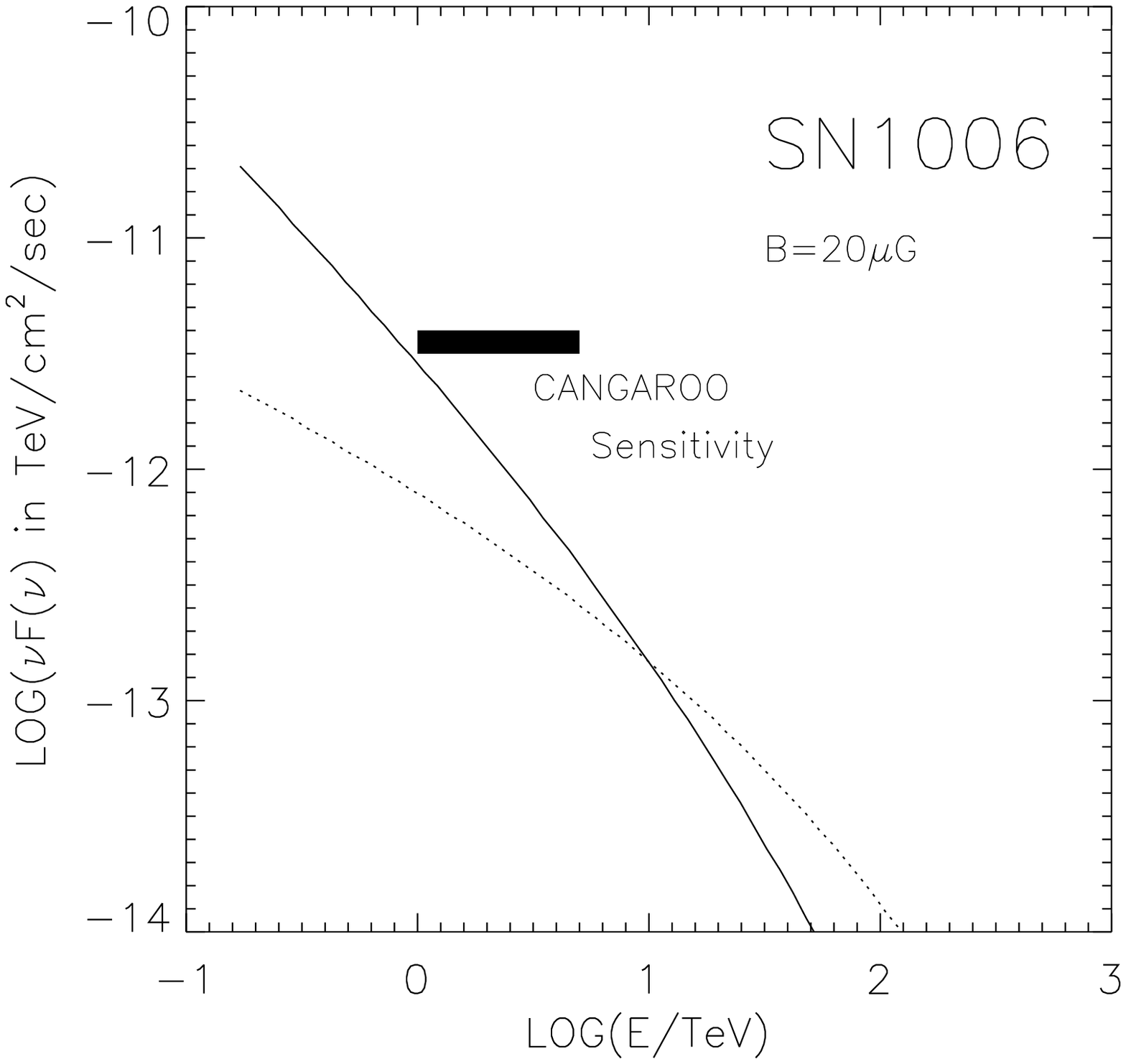,width=8.6cm,clip=}
\figure{1}{The $\gamma$-ray spectrum of SN1006 due to inverse Compton
scattering of microwave photons. The solid line is calculated
for a photon spectral index of s=3 (in the X-rays) while the dotted line is
appropriate for a photon spectral index of s=2.5. The total flux at $\sim$10
TeV, where the curves cross, depends only on the magnetic field strength which
in this plot is taken to be 20 $\mu$G. The dark bar indicates the present
sensitivity of the CANGAROO observatory.}}
%\vskip-11.3cm
%\vskip4.0cm
%\endfig
\vskip0.4truecm
\noindent
Since in our case $\Gamma_e \approx 1$ the Thomson limit is not valid.
The $\gamma$-ray flux at TeV energies is then calculated as
$$I_{\gamma} = {{c \,C (mc^2)^{r-1}}\over {4\pi D^2}} \int_{E_{min}} dE
\int d\epsilon\ n(\epsilon) E^{-r} \sigma(E_\gamma, \epsilon, E)
\ \eqno(6)$$
where
$$n(\epsilon) = {{1}\over {\pi^2 (\hbar c)^3}} {{\epsilon^2}\over 
{\exp\left({{\epsilon}\over {kT}}\right)-1}}\ \eqno(7)$$
is the blackbody spectrum of the microwave background.
The resulting $\nu\,F(\nu )$ flux from SN1006 is given is Fig.1 for the two photon
spectral indices 2.5 and 3. At lower energies the inverse Compton spectrum
has the same power-law behaviour as the synchrotron emission while above
a few TeV it suffers a smooth steepening due to the Klein-Nishina cut-off.
At around 10 TeV, where both curves cross, the flux scales only with
the observed synchrotron flux in X-rays
and with the magnetic field strength which
in this plot is taken to be 20 $\mu G$ according to an estimate on
the basis of the minimum energy assumption. This is in contrast to 
corresponding predictions for the $\gamma$-ray emission of nucleonic origin,
in which both an unknown efficiency factor for the kinetic energy 
transfer from the supernova to cosmic rays and an unknown quantity of 
ambient gas enters the calculation (Drury et al. 1994; Aharonian et al.
1994).

\titlea{Discussion}

The predicted flux can be compared the present sensitivity of TeV
$\gamma$-ray telescope on the southern hemisphere. Air shower arrays
like JANZOS have an energy threshold of around 100 TeV which is beyond the
Klein-Nishina cut-off in the inverse Compton spectra, and hence are
inadequate instruments to detect this emission. However, the current
upper limit for SN1006 ($\le 1.7\cdot 10^{-13}\ {\rm ph./cm^2/sec}$) may
still place constraints on the nucleonic component of cosmic rays in
this object (Allen et al. 1995).

The \v Cerenkov telescope
CANGAROO on the other hand has an energy threshold of around a TeV
which appears advantageous also in view of the particle spectrum
in the remnant of SN1006. We have estimated the current sensitivity
of CANGAROO from the upper limit on another supernova remnant
(Mori et al. 1995). The sensitivity is indicated by the dark bar in Fig.1.
Obviously, the predicted TeV flux for a photon spectral index of 3 and a
mean magnetic field strength of 20$\,\mu G$ nearly meets the sensitivity limit
of CANGAROO. If the spectrum is harder, or if it is gradually rolling over,
the predicted flux is within a factor of 3 of the observing capabilities of
CANGAROO. It is obvious that there must be a roll-over to a flatter spectrum
somewhere between 10 GeV and 10 TeV to harmonize with the EGRET data for
which we derive an upper limit of $2\cdot 10^{-7}\ {\rm ph./cm^2/sec}$
integrated above 100 MeV corresponding to a $\nu F(\nu) = 2\cdot 10^{-11}
\ {\rm TeV/cm^2/sec}$ for a photon spectral index of 2. 

The predicted
integrated photon flux above 1 TeV is
$$I(>1 {\rm TeV})= $$ 
$$\cases{ 1.3\cdot 10^{-12}\  \left({{B}\over {20 \mu G}}\right)^{-3}\ &${\rm cm^{-2} sec^{-1}}$\qquad\ 
for\quad s=3\cr\cr
5\cdot 10^{-13}\  \left({{B}\over {20 \mu G}}\right)^{-2.5}\ &${\rm cm^{-2} sec^{-1}}$\qquad\ 
for\quad s=2.5\cr}\ \eqno(8)$$

The energy loss time scale of an electron with 100 TeV energy in a 
magnetic field of strength 20 $\mu G$ is around 200 years, short enough to 
influence the spectrum considerably. A first estimate of the particle
energy at which synchrotron losses start to alter the spectrum can be derived
by equating the loss time scale and the age of SN1006. The resulting
20 TeV electron energy correspond to synchrotron radiation at 200 eV and to
inverse Compton emission at 1 TeV. It may also be that at 100 TeV the electron
spectrum is not yet in a steady-state. So if the steep synchrotron spectrum 
reflects a smooth cut-off in the electron spectrum due to synchrotron losses counterbalancing the acceleration or also incomplete acceleration,
the expected $\gamma$-ray luminosity
spectrum would be constant at a few times $10^{-12}\ {\rm TeV/cm^2/sec}$
below 1 TeV, more or less follow the dotted line in Fig.1 up to a few TeV,
and then follow the solid line.

\titlea{Conclusions}

In this letter we have calculated the flux of TeV $\gamma$-rays from the
remnant of SN1006 which is expected from inverse Compton scattering
of microwave background photons, if the synchrotron interpretation of the
lineless power-law emission in X-rays is correct. It is shown that
for a reasonable magnetic field strength (20$\,\mu G$) the predicted flux is
of the same order as the present sensitivity limit of the \v Cerenkov
telescope CANGAROO. The TeV $\gamma$-ray emission is a natural consequence
of the synchrotron nature of the X-ray emission and can therefore be used
as a test for the latter interpretation.

The following conclusions can be drawn:

-- With a slight improvement in sensitivity or a lower energy threshold
the CANGAROO telescope should be able to observe inverse Compton scattered
microwave photons
from the remnant of SN1006, if the power-law emission at X-rays is really 
synchrotron radiation. The predicted spectrum is rather steep  with a 
photon spectral index of around 3.

-- If supernova remnants other than SN1006 can be detected by \v Cerenkov
telescopes, this will not provide evidence of shock accelerated protons,
as was proposed by Drury et al. (1994), since substantial emission of
highest energy electrons may not only play a role in SN1006 but also in
other supernova remnants.

-- We strongly encourage high-resolution X-ray spectroscopy observations
of the supernova remnants Cas A and IC443 
which both appear to have power-law
emission spectra above 4 keV (Holt et al. 1994; Wang et al. 1992).

\begref{References}

\ref Aharonian F.A., Drury L.O'C., V\"olk H.J.: 1994, A\&A 285, 645

\ref Allen W.H. and the JANZOS coll.: 1995, 24th ICRC, Rome, OG 4.2.23,
p.447

\ref Ammosov A.E., Ksenofontov L.T., Nikolaev V.S., Petukov S.I.: 1994,
Astr. Lett. 20, 157

\ref Becker R.H., Szymkowiak A.E., Boldt E.A. et al.: 1980, ApJ 240, L33

\ref Blumenthal G.R., Gould R.J.: 1970, Rev. Mod. Phys. 42-2, 237

\ref Drury L.O'C., Aharonian F.A., V\"olk H.J.: 1994, A\&A 287, 959

\ref Holt S.S., Gotthelf E.V., Tsunemi H., Negoro H.: 1994, PASJ 46, L151

\ref Koyama K., Tsunemi H., Becker R.H., Hughes J.P.: 1987, PASJ 39, 437

\ref Koyama K., Petre R., Gotthelf E.V. et al.: 1995, Nat 378, 255

\ref Mori M., Hara T., Hayashida N. et al.: 1995, 24th ICRC, Rome, OG 4.2.33,
p.487

\ref Ozaki M., Koyama K., Ueno S., Yamauchi S.: 1994, PASJ 46, 367

\ref Reynolds S.P., Chevalier R.A.: 1981, ApJ 245, 912

\ref Rybicki G.B., Lightman A.P.: 1979, {\it Radiative Processes in
Astrophysics}, Wiley Interscience

\ref Wang Z.R., Asaoka I., Hayakawa S., Koyama K.: 1992, PASJ 42, 303

\bye
\end